\documentclass[sigconf]{acmart}

\usepackage{balance}

\AtBeginDocument{%
  }

\copyrightyear{2026}
\acmYear{2026}
\setcopyright{cc}
\setcctype{by}
\acmConference[MSR '26]{23rd International Conference on Mining Software Repositories}{April 13--14, 2026}{Rio de Janeiro, Brazil}
\acmBooktitle{23rd International Conference on Mining Software Repositories (MSR '26), April 13--14, 2026, Rio de Janeiro, Brazil}
\acmPrice{}
\acmDOI{10.1145/3793302.3793589}
\acmISBN{979-8-4007-2474-9/2026/04}

\begin{document}

%%
%% The "title" command has an optional parameter,
%% allowing the author to define a "short title" to be used in page headers.
\title{AI IDEs or Autonomous Agents? Measuring the Impact of Coding Agents on Software Development}

%%
%% The "author" command and its associated commands are used to define
%% the authors and their affiliations.
%% Of note is the shared affiliation of the first two authors, and the
%% "authornote" and "authornotemark" commands
%% used to denote shared contribution to the research.

\author{Shyam Agarwal}
\orcid{0009-0009-2147-5674}
\affiliation{%
  \institution{Carnegie Mellon University}
  \city{Pittsburgh}
  \state{Pennsylvania}
  \country{USA}
}
%\email{shyamaga@andrew.cmu.edu}

\author{Hao He}
\orcid{0000-0001-8311-6559}
\affiliation{%
  \institution{Carnegie Mellon University}
  \city{Pittsburgh}
  \state{Pennsylvania}
  \country{USA}
}
%\email{haohe@cmu.edu}

\author{Bogdan Vasilescu}
\orcid{0000-0003-4418-5783}
\affiliation{%
  \institution{Carnegie Mellon University}
  \city{Pittsburgh}
  \state{Pennsylvania}
  \country{USA}
}
%\email{vasilescu@cmu.edu}

%%
%% By default, the full list of authors will be used in the page
%% headers. Often, this list is too long, and will overlap
%% other information printed in the page headers. This command allows
%% the author to define a more concise list
%% of authors' names for this purpose.
\renewcommand{\shortauthors}{Agarwal et al.}

%%
%% The abstract is a short summary of the work to be presented in the
%% article.
\begin{abstract}
  Large language model (LLM) based coding agents increasingly act as autonomous contributors that generate and merge pull requests, yet their real-world effects on software projects are unclear—especially compared with widely adopted IDE-based AI assistants. We present a longitudinal causal study of agent adoption in open-source repositories using staggered difference-in-differences with matched controls. Using the AIDev dataset, we define adoption as the first agent-generated pull request and analyze monthly repository-level outcomes spanning development velocity (commits, lines added) and software quality (static-analysis warnings, cognitive complexity, duplication, and comment density). Results show large, front-loaded velocity gains only when agents are the first observable AI tool in a project; repositories with prior AI IDE usage experience minimal or short-lived throughput increases. In contrast, quality risks are persistent across settings, with static-analysis warnings and cognitive complexity rising by roughly 18\% and 39\%, indicating sustained agent-induced technical debt even when velocity advantages fade. These heterogeneous effects suggest diminishing returns to AI assistance and highlight the need for quality safeguards, provenance tracking, and selective deployment of autonomous agents. Our findings establish an empirical basis for understanding how agentic and IDE-based tools interact, and motivate research on balancing acceleration with maintainability in AI-integrated development workflows. The replication package for this study is publicly available at \url{https://github.com/shyamagarwal13/agentic-coding-impact}.
\end{abstract}

%%
%% The code below is generated by the tool at http://dl.acm.org/ccs.cfm.
%% Please copy and paste the code instead of the example below.
%%


%%
%% Keywords. The author(s) should pick words that accurately describe
%% the work being presented. Separate the keywords with commas.
%% A "teaser" image appears between the author and affiliation
%% information and the body of the document, and typically spans the
%% page.

% \received{20 February 2007}
% \received[revised]{12 March 2009}
% \received[accepted]{5 June 2009}

%%
%% This command processes the author and affiliation and title
%% information and builds the first part of the formatted document.
\maketitle

\section{Introduction}
The landscape of software development is rapidly changing with the growing use of AI-powered coding tools. Importantly, AI-assisted development is not uniform: two distinct paradigms have emerged. The first consists of \textbf{\emph{pre-agentic IDE-based coding assistants}} such as early GitHub Copilot and Cursor, which integrate directly into developers' environments to provide real-time code suggestions and inline assistance \cite{Cheng2022ItWW, Pandey2024TransformingSD, Coutinho2024TheRO}. These tools operate synchronously within the editing workflow, offering suggestions that developers can accept, modify, or reject as they type \cite{Mozannar2022ReadingBT, Guglielmi2025HowDC}.

The second, more recent paradigm involves \textbf{\emph{IDE and web-based coding agents}}—autonomous AI systems that operate at the repository level to generate entire pull requests, implement features, and make substantial code contributions with minimal human intervention. Tools such as OpenAI Codex~\cite{openai-codex}, Anthropic’s Claude Code Agent~\cite{claude-code}, Devin~\cite{devin}, and the Cursor Agent~\cite{cursor} exemplify this emerging category, where AI systems act more like autonomous contributors. Unlike pre-agentic assistants that augment individual coding sessions, coding agents can work asynchronously, potentially completing full features or bug fixes independently \cite{Roychoudhury2025AgenticAF}. They differ from traditional assistants in at least four key dimensions: (1) \emph{Autonomy}—ability to complete tasks without continuous human guidance; (2) \emph{Scope}—operation across multiple files and entire codebases; (3) \emph{Planning}—breaking down requirements into subtasks and executing multi-step solutions; and (4) \emph{Interaction}—primarily contributing through pull requests rather than inline suggestions.

Despite the growing use of agentic coding tools in open-source development, empirical research has largely focused on pre-agentic assistants, in part due to the recency of agentic tools as a technology category. Existing evaluations of coding agents primarily rely on benchmarks without humans-in-the-loop \cite{Becker2025MeasuringTI}, limiting insight into impacts on user experience and productivity \cite{Chen2025CodeWM}. In contrast, pre-agentic coding assistants like GitHub Copilot have been extensively studied, with work documenting positive effects on perceived productivity \cite{Chen2025CodeWM, Vaithilingam2022ExpectationVE, Ziegler2022ProductivityAO}. Research on Copilot spans productivity \cite{Chretien2024ImpactOA, Shihab2025TheEO, Ziegler2024MeasuringGC}, code quality \cite{Nguyen2022AnEE, Yetistiren2022AssessingTQ, Pearce2021AsleepAT}, and developer experience \cite{Liang2023ALS, Weisz2024ExaminingTU}, yet the impacts of autonomous, repository-level agents remain largely unexplored. This gap is concerning given that agentic tools operate with greater autonomy, make larger-scale contributions, and introduce distinct challenges \cite{Sapkota2025VibeCV}.

Understanding the impact of agentic coding tools is critical for several reasons. First, these tools are rapidly gaining traction in open-source development, with an increasing number of repositories accepting AI-generated pull requests \cite{Xiao2025SelfAdmittedGU}. Second, the autonomous nature of agentic tools raises unique questions about code quality, maintainability, and technical debt that may not generalize from IDE-based tool studies \cite{He2025DoesAC}. Third, the scale of contributions, entire features versus individual code snippets, necessitates different evaluation frameworks and metrics \cite{Chen2025CodeWM}. Finally, as organizations consider adopting these tools, evidence-based understanding of their impact on development velocity and software quality becomes essential for informed decision-making \cite{Bridgeford2025TenSR, Ahuchogu2025EvaluatingTI}.

In this paper, we use the AIDev dataset \cite{li2025aidev} to examine how agentic coding tools affect software development velocity and quality at the repository level, reflecting their role as autonomous contributors rather than inline assistants. Focusing on repository-level outcomes, rather than individual-level, captures the fundamentally different operational model of agentic tools. Prior work finds short-term velocity gains but also increased technical debt for repositories adopting the Cursor AI IDE~\cite{He2025DoesAC}. Using similar causal inference methods~\cite{Borusyak2021RevisitingES}, we estimate the effects of adopting a wide range of agentic coding tools, comparing the development trajectories of repositories that used pre-agentic coding tools to those that directly adopted coding agents. Our methodology accounts for staggered adoption, heterogeneous treatment effects, and time-varying confounders, enabling a causal assessment of the marginal impact of autonomous coding agents. Our research addresses:

\begin{itemize}
    \item \textbf{RQ1: Development Velocity Impact} — How does the adoption of coding agents affect development velocity?
    \item \textbf{RQ2: Software Quality Impact} — How does the adoption of coding agents affect software quality?
    \item \textbf{RQ3: Prior AI Exposure and Transition Effects} — How do the effects of adopting coding agents differ between repositories with and without prior IDE-based AI assistance?
\end{itemize}

Our contributions are twofold: (1)~We replicate and extend prior results on newer data and a broader ecosystem of autonomous coding agents, providing the first large-scale longitudinal evidence on agentic contributions at the repository level, and (2)~We provide the first causal evidence on the differential effects of transitioning from IDE-based AI assistants to autonomous coding agents, showing how prior AI tool adoption affects both velocity and code quality.

% (2) an extension of difference-in-differences methodology to study AI agent contributions in open-source development,

\section{Related Work}
Prior work evaluates AI coding tools using development velocity and software quality metrics. Velocity has been measured via task completion time \cite{Paradis2025CreatingBC}, code volume and throughput \cite{Kumar2025IntuitionTE}, and cycle-time reductions \cite{Chretien2024ImpactOA}. Quality outcomes span security vulnerabilities \cite{Pearce2021AsleepAT}, code churn and revision requirements \cite{Xiao2025SelfAdmittedGU}, static analysis warnings and code complexity \cite{He2025DoesAC}, and broader indicators of technical debt. Methodologically, studies employ randomized trials \cite{Paradis2025CreatingBC}, observational field studies \cite{Chretien2024ImpactOA}, and quasi-experimental designs including difference-in-differences \cite{He2025DoesAC}.

The vast majority of real-world research focuses on IDE-based AI assistants \cite{Borek2025QualityEO, Chretien2024ImpactOA, Paradis2024HowMD, Ng2024HarnessingTP, Shihab2025TheEO, Taivalsaari2025OnTF, Nguyen2022AnEE, Yetistiren2022AssessingTQ, Pearce2021AsleepAT, Asare2022IsGC, Xiao2025SelfAdmittedGU, Xu2025AIassistedPM}, typically reporting modest velocity improvements \cite{He2025DoesAC, Vaithilingam2022ExpectationVE, Ziegler2022ProductivityAO, Imai2022IsGC}. In contrast, early studies of agentic tools show mixed results: maintainers merge the majority of Claude Code pull requests (83.8\% of 567) \cite{watanabe2025use}, yet controlled experiments with tools such as Cursor find limited productivity benefits for experienced open-source developers \cite{Becker2025MeasuringTI}, potentially due to over-optimism, unreliability, and high task complexity. These inconsistencies underscore the need for longitudinal evidence on the real-world effects of autonomous coding agents.

\section{Methods}

We estimate the causal effects of adopting \emph{LLM-based coding agents} on project-level development velocity and software quality using a quasi-experimental difference-in-differences (DiD) design with staggered adoption. Our empirical strategy, outcome definitions, matching procedure, and estimation methods follow similar work on Cursor adoption~\cite{He2025DoesAC}, unless otherwise specified, enabling direct comparability. We summarize the methodology and highlight deviations specific to coding agents.

\subsection{Data Collection and Treatment Definition}

\looseness=-1
We build on the AIDev dataset (v3) \cite{li2025aidev}, which links GitHub repositories to AI-generated pull requests and provides explicit evidence of agentic contributions (e.g., multi-file edits, autonomous refactorings, test and documentation generation). For each repository, we define the \emph{agent adoption date} as the earliest month containing an agent-attributed PR. Because AIDev coverage begins in December 2024 while most agentic tools were released earlier that year, we mitigate left-censoring by retrospectively parsing all PRs from January 2024 through November 2025, ensuring that the earliest observed agentic PR closely approximates true adoption timing—crucial for valid temporal ordering in our staggered DiD design.

\begin{table}[t]
\centering
\scriptsize
\caption{Descriptive statistics of treated repositories by prior AI exposure. AF repositories are older but smaller and less popular, while IF repositories are more starred, forked, and active; both exhibit substantial agentic contributions.}
\label{tab:repo-stats}
\begin{tabular}{lcccc}
\toprule
Outcome & Mean (AF/IF) & Min (AF/IF) & Median (AF/IF) & Max (AF/IF) \\
\midrule
Age (days) \textsuperscript{\textdagger}           & 1537.8/1215.2 & 131/199   & 1069/926   & 6141/5570 \\
Stars \textsuperscript{\textdagger\textdagger}                & 1460.9/8123.4 & 10/10     & 68/423     & 48110/177379 \\
Forks \textsuperscript{\textdagger\textdagger}                & 236.8/1340.2  & 0/0       & 12/120      & 8940/45908 \\
Pull Requests \textsuperscript{\textdagger}        & 696.7/2146.0  & 10/21     & 122/740    & 31836/35389 \\
Agentic Pull Requests \textsuperscript{\textdagger}  & 121.5/101.5    & 10/10     & 28/37      & 13462/912 \\
\bottomrule
\end{tabular}
\raggedright
\scriptsize \textit{Note:} 
$\dagger$ as of November end;\;
$\dagger\dagger$ as in original dataset.\;
\end{table}

\begin{figure}
    \centering
    \includegraphics[width=\linewidth]{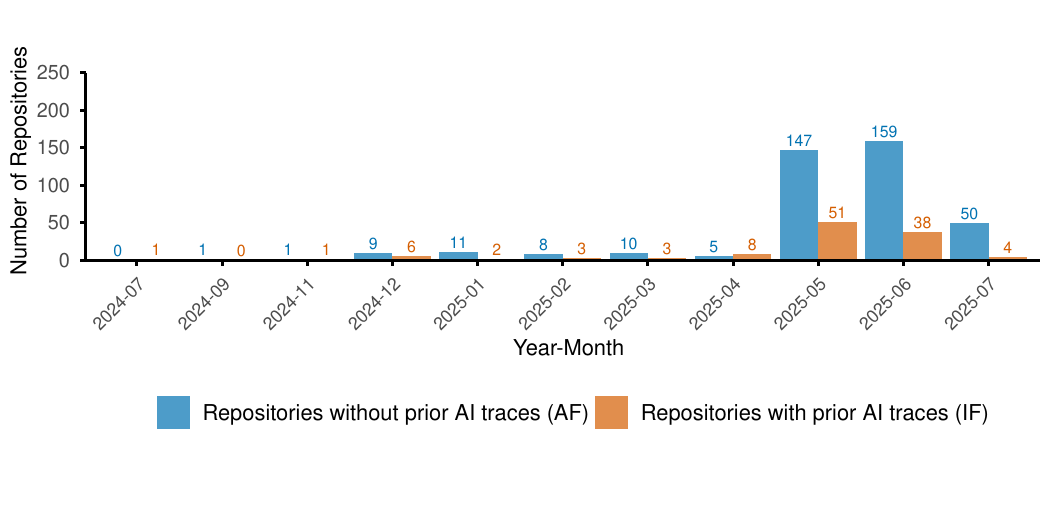}
    \caption{Monthly distribution of agent adoption dates, separated by prior AI exposure. Most adoptions occur between May--July 2025.
    % , with AF repositories more prevalent overall.
    }
    \label{fig:time-agent-adoption}
\end{figure}

We extend the original agent taxonomy (Claude \cite{claude-code}, Copilot \cite{copilot}, Cursor \cite{cursor}, Devin \cite{devin}, Codex \cite{openai-codex}) to mutually exclusive labels: \emph{human}, \emph{bot}, \emph{codex} \cite{openai-codex}, \emph{devin} \cite{devin}, \emph{jules} \cite{jules}, \emph{cursor} \cite{cursor}, \emph{claude} \cite{claude-code}, \emph{copilot} \cite{copilot}, \emph{openhands} \cite{openhands}, \emph{codegen} \cite{codegen}, \emph{cosine} \cite{cosine}, and \emph{tembo} \cite{tembo}. Attribution follows a cascading strategy prioritizing agent-specific signals before generic labels: (1) branch prefixes (e.g., \emph{cursor/}, \emph{claude/}); (2) PR author logins (e.g., \emph{codegen-sh}, \emph{tembo-io}); (3) first-commit author names (e.g., \emph{google-labs-jules[bot]}, \emph{claude[bot]}); (4) GitHub actor type \emph{bot}; and (5) default classification as \emph{human}. For Claude-specific attribution, we additionally search PR descriptions, comments, and reviews for distinctive co-authorship patterns (“Generated with Claude Code”, optionally including “Co-Authored-By: Claude <no-reply@anthropic.com>”). This multi-signal approach improves recall for agentic PRs lacking explicit bot authorship or branch naming.

By following this convention, we identified several misclassifications and missing PRs in the original dataset (replication package). Correctly recovering the \emph{earliest} agentic PR is essential because adoption timing determines treatment assignment in our causal design; any remaining attribution errors primarily introduce noise in treatment timing rather than systematic bias, likely attenuating effects toward zero.

We restrict analysis to repositories with $\geq$10 stars to exclude toy and spam projects and require $\geq$10 agentic PRs per treated repository to ensure non-trivial exposure. Repositories are observed monthly before and after adoption, forming an unbalanced panel with staggered timing. To examine moderation by prior AI usage, we partition treated repositories into \emph{agent-first} (AF), which show no traces of AI IDEs throughout the collection time period, and \emph{IDE-first} (IF), which exhibit IDE activity prior to agent adoption. Prior IDE usage is detected monthly by scanning the most recent commit for configuration artifacts associated with agent-enabled IDEs (GitHub Copilot~\cite{copilot}, Cursor~\cite{cursor}, WindSurf~\cite{windsurf}). A repository is IF if such artifacts appear before its first agentic PR and AF if they don't exist at all. Analyses are conducted separately for AF and IF using control repositories that mirror the same IDE exposure pattern. Figure~\ref{fig:time-agent-adoption} shows both groups adopt primarily between April--June~2025, with AF more numerous. Table~\ref{tab:repo-stats} shows AF repos are older but smaller and less popular, whereas IF repos are more starred, forked, and PR-active; both groups nonetheless exhibit substantial agentic PR volume, motivating separate analyses.

Our control set consists of all GitHub repositories with $\geq$10 stars at collection time. For each month with agent adoption, we extract monthly activity series from GHArchive~\cite{gharchive} for repositories with at least one event (age, active users, stars, forks, releases, pull requests, issues, comments, total events). Because inferring prior IDE exposure at this scale is initially infeasible, we first perform propensity-score matching over the full control population without conditioning on prior exposure. We then infer prior exposure only for matched candidates and retain matched control sets that mirror treated repositories’ AF/IF status. After filtering, our heterogeneous samples contain 401 AF repositories matched to 606 controls and 117 IF repositories matched to 73 controls.

% \begin{table*}
% \small
% \centering
% \caption{Descriptive statistics of treated repositories partitioned by prior AI exposure: AF (agent-first) and IF (IDE-first). AF repositories are generally older but smaller and less popular, while IF repositories have more stars, forks, and pull requests; both groups exhibit substantial agentic contribution volumes.}
% \label{tab:repo-stats}
% \begin{threeparttable}
% \begin{tabular}{l
%   r|l r|l r|l r|l r|l r|l}
% \toprule
%  & \multicolumn{2}{c|}{Mean} 
%  & \multicolumn{2}{c|}{Min} 
%  & \multicolumn{2}{c|}{25\%} 
%  & \multicolumn{2}{c|}{Median} 
%  & \multicolumn{2}{c|}{75\%} 
%  & \multicolumn{2}{c}{Max} \\
% \cmidrule(lr){2-3}
% \cmidrule(lr){4-5}
% \cmidrule(lr){6-7}
% \cmidrule(lr){8-9}
% \cmidrule(lr){10-11}
% \cmidrule(lr){12-13}
%  & AF & IF & AF & IF & AF & IF & AF & IF & AF & IF & AF & IF \\
% \midrule
% Age (days)             & 1567.8 & 1268.9 & 161 & 197 & 413.5 & 427 & 1099 & 955 & 2345 & 1761.5 & 6171 & 5600 \\
% Stars                  & 1460.9 & 7531.3 & 10 & 10 & 22 & 53.5 & 68 & 367 & 320 & 4406.5 & 48110 & 177379 \\
% Forks                  & 236.8 & 1143.3 & 0 & 0 & 4 & 13.5 & 12 & 90 & 61 & 581 & 8940 & 45908 \\
% Pull Requests          & 696.7 & 1978.4 & 10 & 16 & 40 & 257.5 & 122 & 693 & 439 & 2079 & 31836 & 35389 \\
% Agentic Pull Requests  & 121.5 & 93.8 & 10 & 10 & 15 & 21 & 28 & 37 & 62 & 79 & 13462 & 912 \\
% \bottomrule
% \end{tabular}
% \end{threeparttable}
% \end{table*}

\begin{figure*}[!t]
    \centering
    \includegraphics[width=\linewidth]{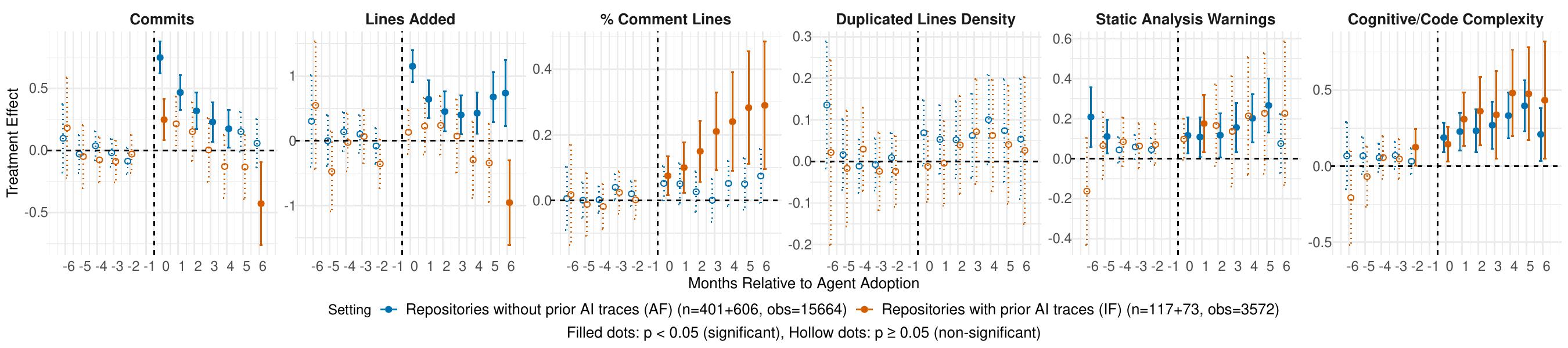}
    \caption{Estimated post-adoption effects of agentic coding tools by prior AI exposure. AF repositories gain velocity and accumulate maintainability risks; IF repositories show minimal velocity gains but comparable maintainability increases.}
    \label{fig:dynamic_effects}
\end{figure*}

\subsection{Outcomes and Covariates}
We measure development velocity via monthly commit counts and lines added; software quality via static-analysis warnings, duplicated-line density, and cognitive complexity using SonarQube\cite{sonarqube}. All outcomes are aggregated at the repository–month level. We control for time-varying covariates: lines of code, repository age (days), contributors, stars received, issues opened, and issue comments added in the observation month. Lines of code nd outcome metrics come from SonarQube~\cite{sonarqube}, contributor counts from version history, and activity covariates from GHArchive~\cite{gharchive}. These covariates are collected consistently for both treated and control repositories across the full observation window, ensuring comparable temporal coverage and supporting adjustment for confounding dynamics unrelated to agent adoption.

\subsection{Causal Inference Strategy}
We estimate effects using DiD with staggered adoption: later adopters serve as controls for earlier adopters prior to adoption; never-adopters provide additional counterfactuals. To address selection bias of control repositories and satisfy the conditional independence assumption to establish the quasi-experimental settings, we perform propensity score matching prior to estimation. Propensity scores use logistic regression over dynamic pre-treatment characteristics, incorporating both activity \textit{levels} (recent covariates) and \textit{trajectories} (lagged covariates) to capture whether repositories are growing, declining, or stable. Specifically, we model adoption likelihood using repository age at the month prior to observation, six monthly covariate lags, and cumulative historical covariates, enabling discrimination between repositories with similar recent activity but different long-term trends. Because candidates outnumber adopters, we subsample at most 10{,}000 candidates/month, yielding propensity models with AUC~0.92–0.99. Treated repositories are matched to maximum three controls with similar propensity scores and the same primary language, ensuring comparable pre-adoption activity, exposure histories, and language-specific performance characteristics. We estimate average and dynamic treatment effects using the imputation-based DiD estimator of Borusyak et al.~\cite{Borusyak2021RevisitingES}, which avoids biases from traditional two-way fixed effects models under staggered adoption and heterogeneous effects. Event-study specifications are used to assess pre-treatment trends and to characterize how effects evolve over time. Standard errors are clustered at the repository level. As in prior work, our estimates should be interpreted as intent-to-treat effects of observable agent adoption. While we cannot directly measure usage intensity or developer-level interactions, staggered adoption and matching on pre-treatment dynamics strengthen the causal interpretation of our findings.

\section{Results and Discussion}

% \begin{table*}[t]
% \centering
% \small
% \caption{The \citet{Borusyak2021RevisitingES} estimated average treatment effects post Agent adoption. All outcome variables are log-transformed to facilitate comparison across outcomes. Treatment effects $\beta$ can be interpreted as percentage change $100(e^\beta-1)\%$.}
% \label{tab:average-te}
% \begin{threeparttable}
% \begin{tabular}{l
%         r|l
%         r|l
%         r|l}
% \toprule
%  & \multicolumn{2}{c}{Estimate $\beta$}
%  & \multicolumn{2}{c}{Std. Error}
%  & \multicolumn{2}{c}{\% Change} \\
% \cmidrule(lr){2-3}
% \cmidrule(lr){4-5}
% \cmidrule(lr){6-7}
% Outcome & AF & IF & AF & IF & AF & IF \\
% \midrule
% Commits & 0.309$^{***}$ & 0.039\phantom{$^{***}$} &
% (0.051) & (0.088) &
% 36.254\% & 4.010\% \\
% Lines Added & 0.569$^{***}$ & 0.010\phantom{$^{***}$} &
% (0.103) & (0.171) &
% 76.587\% & 1.005\% \\

% Duplicate Line Density & 0.076$^{*}$\phantom{$^{**}$} & -0.075\phantom{$^{***}$} &
% (0.044) & (0.050) &
% 7.922\% & -7.240\% \\

% Comment Line Density & 0.042\phantom{$^{***}$} & 0.175$^{***}$ &
% (0.028) & (0.052) &
% 4.338\% & 19.125\% \\

% Static Analysis Warnings & 0.163$^{***}$ & 0.159\phantom{$^{***}$} &
% (0.048) & (0.099) &
% 17.731\% & 17.254\% \\

% Code Complexity & 0.299$^{***}$ & 0.302$^{**}$\phantom{$^{*}$} &
% (0.059) & (0.100) &
% 34.854\% & 35.207\% \\

% \bottomrule
% \multicolumn{7}{r}{\textit{Note:} $^{*}p<0.05$;\;\;$^{**}p<0.01$;\;\;$^{***}p<0.001$}
% \end{tabular}
% \end{threeparttable}
% \end{table*}
 Table~\ref{tab:average-te} summarizes average post-adoption effects, while Figure~\ref{fig:dynamic_effects} shows six-month dynamic estimates.

\begin{table}[t]
\centering
\scriptsize
\caption{The \citet{Borusyak2021RevisitingES} estimated average post-adoption treatment effects of agentic coding tools separated by prior AI exposure. The estimate and standard error are log-transformed to facilitate easy comparison.}
\label{tab:average-te}
\begin{tabular}{lccc}
\toprule
Outcome & $\beta$ (AF / IF) & Std. Err. (AF / IF) & \% Change (AF / IF) \\
\midrule
Commits                & 0.309$^{***}$/0.030       & (0.051)/(0.092) & 36.25 / 3.06 \\
Lines Added            & 0.569$^{***}$/--0.066       & (0.103)/(0.189) & 76.59 / --6.34 \\
Duplicate Line Density & 0.076/--0.009       & (0.044)/(0.056) & 7.92 / --0.94 \\
Comment Line Density   & 0.042/0.201$^{***}$       & (0.028)/(0.055) & 4.34 / 22.30 \\
Static Analysis Warnings & 0.163$^{***}$/0.174     & (0.048)/(0.114) & 17.73 / 19.00 \\
Code Complexity        & 0.299$^{***}$/0.357$^{**}$ & (0.059)/(0.114) & 34.85 / 42.87 \\
\bottomrule
\end{tabular}

\raggedright
\scriptsize \textit{Note:} $^{*}p<0.05$;\;\;$^{**}p<0.01$;\;\;$^{***}p<0.001$.
\end{table}

\smallskip\noindent\textbf{Development Velocity.}
Agentic tools substantially accelerate development only when introduced as a repository’s first observable AI tool. AF repos see large average gains ($+36.3\%$ commits; $+76.6\%$ lines added), whereas IF repos show minimal changes ($+3.1\%$; $-6.3\%$). Dynamics reveal a sharper contrast: AF repositories experience a spike at $t{=}0$ (about $+111\%$ commits; $+216\%$ lines added) and persistently elevated activity, with lines added remaining roughly $+49$–$+109\%$ through $t{=}6$. IF repositories show only a short-lived bump around adoption (commits $+16$–$28\%$ at $t{=}0$–$2$) before estimates return near zero and eventually turn negative (lines ${\sim}-61\%$, commits ${\sim}-35\%$ by $t{=}6$).

These patterns indicate that agentic tools act as high-throughput contributors primarily in new-to-AI workflows, but yield diminishing returns in AI-saturated ones. AF repositories appear able to harvest ``first AI'' acceleration, while IF repositories, having already absorbed productivity gains from AI IDEs, likely face higher coordination and integration costs that limit throughput. The greater maturity of IF repos (higher stars, forks, PR volume; from Table~\ref{tab:repo-stats}) likely constrains how aggressively agentic changes can be merged, so that localized speedups are offset by triage and review overhead. Overall, agent adoption yields sustained velocity gains only when not preceded by IDE-based AI tooling.

\smallskip\noindent\textbf{Software Quality.}
Regardless of prior AI exposure, adoption is associated with increased maintainability risks. Across both AF and IF repositories, static-analysis warnings rise by about $18\%$ and cognitive complexity by roughly $39\%$ (Table~\ref{tab:average-te}). Dynamics show persistent complexity accumulation: AF repos increase $+20.7\%$ at $t{=}0$ and reach ${\sim}+49\%$ by $t{=}5$, while IF repositories become significantly elevated as early as $t{=}{-}2$ and remain ${\sim}+15$–$+62\%$ through $t{=}6$. Warning growth is somewhat noisier but similarly persistent ($\sim+22$–$+31\%$ in AF by $t{=}4$–$5$; IF values consistently positive around $+25\%$ at $t{=}4$–$6$).

These simultaneous increases in complexity and warnings, even when net velocity gains are weak or negative (IF), indicate \emph{agent-induced complexity debt}: agents accelerate the introduction of code that raises long-term cognitive and maintenance load. Duplication effects are small and inconsistent, suggesting quality risks stem from structural complexity rather than copy–paste proliferation. Comment density diverges across groups: IF repositories show substantial and sustained increases (about $+22\%$ on average and ${>}+30\%$ by $t{=}6$), whereas AF effects are muted, hinting that teams already using AI IDEs rely on agents for documentation as well as code. While additional comments can aid comprehension, they do not counterbalance persistent complexity growth. Relative to prior work on AI IDEs~\cite{He2025DoesAC}, which finds modest productivity gains and mixed quality effects, these findings suggest autonomous agents amplify the speed–maintainability trade-off, and in AI-rich environments may magnify complexity without delivering sustained velocity benefits. We also observe isolated significant pre-treatment coefficients in static-analysis warnings and code complexity, suggesting that untreated potential outcomes are not fully captured by additive fixed effects, reflecting systematic mean differences between treated and matched controls. These are not pervasive enough to indicate a sustained pre-trend or clear violation of parallel trends, but they are concerning and highlight a limitation of our quasi-experimental design and the underlying data.

\smallskip\noindent\textbf{Implications and Ethical Considerations}
\label{sec:implications}
Autonomous coding agents function as powerful but risky accelerators whose value depends on prior AI exposure: in our sample substantial front-loaded velocity gains materialize only when agents are a project’s first AI tool; more broadly, velocity increases likely won't suffice, and AI adoption will need to be paired with strong quality safeguards (e.g., complexity-aware review of agent pull requests, routine refactoring, comprehensive automated tests) to prevent accumulating complexity debt. Teams using AI IDEs should not assume additive productivity and may instead deploy agents selectively or for tightly scoped tasks. Persistent complexity growth underscores the need to surface maintainability metrics directly in agent planning and prompting, while modulating behavior according to existing AI usage. Ethically, increased automation shifts accountability and maintainability burdens; provenance tracking, transparency of agent-generated changes, and review practices that emphasize human oversight are still essential to avoid debt.

\section{Conclusion}
Autonomous agents offer meaningful velocity gains only in new-to-AI settings while consistently raising complexity and warning levels across contexts, reinforcing a speed–maintainability trade-off. Prior exposure to AI IDEs moderates benefits but not risks, underscoring the need for selective deployment and active oversight. Future work should follow post-adoption trajectories over longer horizons and examine collaborative patterns that balance acceleration with sustained code quality.

\newpage
\balance
\bibliographystyle{ACM-Reference-Format}
\bibliography{main}

@String{Computing = "Computing" }

@article{Ziegler2022ProductivityAO,
  title={Productivity assessment of neural code completion},
  author={Albert Ziegler and Eirini Kalliamvakou and Shawn Simister and Ganesh Sittampalam and Alice Li and Andrew Rice and Devon Rifkin and Edward Aftandilian},
  journal={Proceedings of the 6th ACM SIGPLAN International Symposium on Machine Programming},
  year={2022},
}

@article{Chretien2024ImpactOA,
  title={Impact of AI-tooling on the Engineering Workspace},
  author={Lena Chretien and Nikolas Albarran},
  journal={ArXiv},
  year={2024},
  volume={abs/2406.07683},
}

@inproceedings{He2025DoesAC,
  title={Speed at the Cost of Quality: How Cursor AI Increases Short-Term Velocity and Long-Term Complexity in Open-Source Projects},
  author={Hao He and Courtney Miller and Shyam Agarwal and Christian Kastner and Bogdan Vasilescu},
  booktitle={International Conference on Mining Software Repositories (MSR)},
  year={2026},
}

@article{li2025aidev,
title={{The Rise of AI Teammates in Software Engineering (SE) 3.0: How Autonomous Coding Agents Are Reshaping Software Engineering}}, 
author={Li, Hao and Zhang, Haoxiang and Hassan, Ahmed E.},
journal={arXiv preprint arXiv:2507.15003},
year={2025}
}

@inproceedings{Borusyak2021RevisitingES,
  title={Revisiting event study designs: robust and efficient estimation},
  author={Kirill Borusyak and Xavier Jaravel and Jann Spiess},
  year={2021},
}

@article{Paradis2025CreatingBC,
  title={Creating benchmarkable components to measure the quality of AI-enhanced developer tools},
  author={Elise Paradis and Ambar Murillo and Maulishree Pandey and Sarah D’Angelo and Andrew Macvean and Ben Ferrari-Church and Matthew Hughes},
  journal={Proceedings of the Extended Abstracts of the CHI Conference on Human Factors in Computing Systems},
  year={2025},
}

@article{Pearce2021AsleepAT,
  title={Asleep at the Keyboard? Assessing the Security of GitHub Copilot’s Code Contributions},
  author={Hammond A. Pearce and Baleegh Ahmad and Benjamin Tan and Brendan Dolan-Gavitt and Ramesh Karri},
  journal={2022 IEEE Symposium on Security and Privacy (SP)},
  year={2021},
  pages={754-768},
}

@article{Xiao2025SelfAdmittedGU,
  title={Self-Admitted GenAI Usage in Open-Source Software},
  author={Tao Xiao and Youmei Fan and Fabio Calefato and Christoph Treude and Raula Gaikovina Kula and Hideaki Hata and Sebastian Baltes},
  journal={ArXiv},
  year={2025},
  volume={abs/2507.10422},
}

@article{Becker2025MeasuringTI,
  title={Measuring the Impact of Early-2025 AI on Experienced Open-Source Developer Productivity},
  author={Joel Becker and Nate Rush and Elizabeth Barnes and David Rein},
  journal={ArXiv},
  year={2025},
  volume={abs/2507.09089},
}

@article{Kumar2025IntuitionTE,
  title={Intuition to Evidence: Measuring AI's True Impact on Developer Productivity},
  author={Anand Kumar and Vishal Khare and Deepak Sharma and Satyam Kumar and Vijay Saini and Anshul Yadav and Sachendra Jain and Ankit Rana and Pratham Verma and Vaibhav Meena and Avinash Edubilli},
  journal={ArXiv},
  year={2025},
  volume={abs/2509.19708},
}

@article{Chen2025CodeWM,
  title={Code with Me or for Me? How Increasing AI Automation Transforms Developer Workflows},
  author={Valerie Chen and Ameet Talwalkar and Robert Brennan and Graham Neubig},
  journal={ArXiv},
  year={2025},
  volume={abs/2507.08149},
}

@article{Vaithilingam2022ExpectationVE,
  title={Expectation vs. Experience: Evaluating the Usability of Code Generation Tools Powered by Large Language Models},
  author={Priyan Vaithilingam and Tianyi Zhang and Elena L. Glassman},
  journal={CHI Conference on Human Factors in Computing Systems Extended Abstracts},
  year={2022},
}

@article{Imai2022IsGC,
  title={Is GitHub Copilot a Substitute for Human Pair-programming? An Empirical Study},
  author={Saki Imai},
  journal={2022 IEEE/ACM 44th International Conference on Software Engineering: Companion Proceedings (ICSE-Companion)},
  year={2022},
  pages={319-321},
}

@article{Shihab2025TheEO,
  title={The Effects of GitHub Copilot on Computing Students' Programming Effectiveness, Efficiency, and Processes in Brownfield Coding Tasks},
  author={Md. Istiak Hossain Shihab and Christopher Hundhausen and Ahsun Tariq and Summit Haque and Yunhan Qiao and Brian Mulanda},
  journal={Proceedings of the 2025 ACM Conference on International Computing Education Research V.1},
  year={2025},
}

@article{Ziegler2024MeasuringGC,
  title={Measuring GitHub Copilot's Impact on Productivity},
  author={Albert Ziegler and Eirini Kalliamvakou and LI X.ALICE and Andrew Rice and Devon Rifkin and Shawn Simister and Ganesh Sittampalam and Edward Aftandilian},
  journal={Communications of the ACM},
  year={2024},
  volume={67},
  pages={54 - 63},
}

@article{Yetistiren2022AssessingTQ,
  title={Assessing the quality of GitHub copilot’s code generation},
  author={Burak Yetistiren and Isik Ozsoy and Eray T{\"u}z{\"u}n},
  journal={Proceedings of the 18th International Conference on Predictive Models and Data Analytics in Software Engineering},
  year={2022},
}

@article{Nguyen2022AnEE,
  title={An Empirical Evaluation of GitHub Copilot's Code Suggestions},
  author={Nhan Ton Nguyen and Sarah Nadi},
  journal={2022 IEEE/ACM 19th International Conference on Mining Software Repositories (MSR)},
  year={2022},
  pages={1-5},
}

@article{Liang2023ALS,
  title={A Large-Scale Survey on the Usability of AI Programming Assistants: Successes and Challenges},
  author={Jenny T Liang and Chenyang Yang and Brad A. Myers},
  journal={2024 IEEE/ACM 46th International Conference on Software Engineering (ICSE)},
  year={2023},
  pages={616-628},
}

@article{Weisz2024ExaminingTU,
  title={Examining the Use and Impact of an AI Code Assistant on Developer Productivity and Experience in the Enterprise},
  author={Justin D. Weisz and Shraddha Kumar and Michael J. Muller and Karen-Ellen Browne and Arielle Goldberg and Katrin Ellice Heintze and Shagun Bajpai},
  journal={Proceedings of the Extended Abstracts of the CHI Conference on Human Factors in Computing Systems},
  year={2024},
}

@article{Coutinho2024TheRO,
  title={The Role of Generative AI in Software Development Productivity: A Pilot Case Study},
  author={Mariana Coutinho and Lorena Marques and Anderson Santos and Marcio Dahia and C{\'e}sar França and Ronnie de Souza Santos},
  journal={Proceedings of the 1st ACM International Conference on AI-Powered Software},
  year={2024},
}

@article{Bridgeford2025TenSR,
  title={Ten Simple Rules for AI-Assisted Coding in Science},
  author={Eric W. Bridgeford and Iain Campbell and Zijao Chen and Zhicheng Lin and Harrison Ritz and Joachim Vandekerckhove and R.A. Poldrack},
  journal={ArXiv},
  year={2025},
  volume={abs/2510.22254},
}

@article{Ahuchogu2025EvaluatingTI,
  title={Evaluating the Impact of Generative AI on Intelligent Programming Assistance and Code Quality.},
  author={Magnus Chukwuebuka Ahuchogu and Pravin Ganpatrao Gawande and Dr Charu Mohla and Dr. Deepak A. Vidhate and Nidal Al Said},
  journal={Power System Technology},
  year={2025},
}

@online{gharchive,
  year={2011},
  title={{GHArchive}},
  url={https://www.gharchive.org/},
  lastaccessed={Nov 30, 2024}
}

@article{Xu2025AIassistedPM,
  title={AI-assisted Programming May Decrease the Productivity of Experienced Developers by Increasing Maintenance Burden},
  author={Feiyang Xu and Poonacha K. Medappa and Murat Mustafa Tunç and Martijn Vroegindeweij and Jan C Fransoo},
  journal={ArXiv},
  year={2025},
  volume={abs/2510.10165},
}

@article{Asare2022IsGC,
  title={Is GitHub’s Copilot as bad as humans at introducing vulnerabilities in code?},
  author={Owura Asare and Meiyappan Nagappan and Nirmal Asokan},
  journal={Empirical Software Engineering},
  year={2022},
  volume={28},
  pages={1-24},
}

@article{Taivalsaari2025OnTF,
  title={On the Future of Software Reuse in the Era of AI Native Software Engineering},
  author={Antero Taivalsaari and Tommi Mikkonen and Cesare Pautasso},
  journal={ArXiv},
  year={2025},
  volume={abs/2508.19834},
}

@article{Ng2024HarnessingTP,
  title={Harnessing the Potential of Gen-AI Coding Assistants in Public Sector Software Development},
  author={Kevin KB Ng and Liyana Fauzi and Leon Leow and Jaren Ng},
  journal={ArXiv},
  year={2024},
  volume={abs/2409.17434},
}

@article{Paradis2024HowMD,
  title={How Much Does AI Impact Development Speed? an Enterprise-Based Randomized Controlled Trial},
  author={Elise Paradis and Kate Grey and Quinn Madison and Daye Nam and Andrew Macvean and Vahid Meimand and Nan Zhang and Ben Ferrari-Church and Satish Chandra},
  journal={2025 IEEE/ACM 47th International Conference on Software Engineering: Software Engineering in Practice (ICSE-SEIP)},
  year={2024},
  pages={618-629},
}

@article{Borek2025QualityEO,
  title={Quality evaluation of Tabby coding assistant using real source code snippets},
  author={Marta Borek and Robert Nowak},
  journal={ArXiv},
  year={2025},
  volume={abs/2504.08650},
}

@article{watanabe2025use,
  title={On the use of agentic coding: An empirical study of pull requests on GitHub},
  author={Watanabe, Miku and Li, Hao and Kashiwa, Yutaro and Reid, Brittany and Iida, Hajimu and Hassan, Ahmed E},
  journal      = {CoRR},
  volume       = {abs/2509.14745},
  year={2025}
}

@online{sonarqube,
  year={2025},
  title={SonarQube Community Build Documentation},
  url={https://docs.sonarsource.com/sonarqube-community-build/},
  lastaccessed={Nov 30, 2025}
}

@online{windsurf,
  year         = {2025},
  title        = {Windsurf - The best AI for Coding},
  url          = {https://windsurf.com},
  lastaccessed = {Nov 30, 2025}
}

@online{openhands,
  year         = {2025},
  title        = {OpenHands},
  url          = {https://app.all-hands.dev/},
  lastaccessed = {Nov 30, 2025}
}

@online{cursor,
  year         = {2025},
  title        = {Cursor - The AI Code Editor},
  url          = {https://www.cursor.com/},
  lastaccessed = {Nov 30, 2025}
}

@online{openai-codex,
  year         = {2025},
  title        = {Codex | OpenAI},
  url          = {https://openai.com/codex/},
  lastaccessed = {Nov 30, 2025}
}

@online{claude-code,
  year         = {2025},
  title        = {Claude Code | Claude},
  url          = {https://claude.com/product/claude-code},
  lastaccessed = {Nov 30, 2025}
}

@online{devin,
  year         = {2025},
  title        = {Devin | The AI Software Engineer},
  url          = {https://devin.ai/},
  lastaccessed = {Nov 30, 2025}
}

@online{copilot,
  year         = {2025},
  title        = {GitHub Copilot · Your AI pair programmer},
  url          = {https://github.com/features/copilot},
  lastaccessed = {Nov 30, 2025}
}

@online{jules,
  year         = {2025},
  title        = {Jules - An Autonomous Coding Agent},
  url          = {https://jules.google/},
  lastaccessed = {Nov 30, 2025}
}

@online{codegen,
  year         = {2025},
  title        = {Codegen | The OS for Code Agents},
  url          = {https://codegen.com/},
  lastaccessed = {Nov 30, 2025}
}

@online{cosine,
  year         = {2025},
  title        = {Self sufficient agentic AI software engineer | Cosine AI},
  url          = {https://cosine.sh/},
  lastaccessed = {Nov 30, 2025}
}

@online{tembo,
  year         = {2025},
  title        = {Tembo — Delegate work to any coding agent},
  url          = {https://www.tembo.io/},
  lastaccessed = {Nov 30, 2025}
}

@article{Cheng2022ItWW,
  title={“It would work for me too”: How Online Communities Shape Software Developers’ Trust in AI-Powered Code Generation Tools},
  author={Ruijia Cheng and Ruotong Wang and Thomas Zimmermann and Denae Ford},
  journal={ACM Transactions on Interactive Intelligent Systems},
  year={2022},
  volume={14},
  pages={1 - 39},
}

@article{Pandey2024TransformingSD,
  title={Transforming Software Development: Evaluating the Efficiency and Challenges of GitHub Copilot in Real-World Projects},
  author={Ruchika Pandey and Prabhat Singh and Raymond Wei and Shaila Shankar},
  journal={ArXiv},
  year={2024},
  volume={abs/2406.17910},
}

@article{Mozannar2022ReadingBT,
  title={Reading Between the Lines: Modeling User Behavior and Costs in AI-Assisted Programming},
  author={Hussein Mozannar and Gagan Bansal and Adam Fourney and Eric Horvitz},
  journal={ArXiv},
  year={2022},
  volume={abs/2210.14306},
}

@article{Guglielmi2025HowDC,
  title={How do Copilot Suggestions Impact Developers' Frustration and Productivity?},
  author={Emanuela Guglielmi and Venera Arnoudova and Gabriele Bavota and Rocco Oliveto and Simone Scalabrino},
  journal={ArXiv},
  year={2025},
  volume={abs/2504.06808},
}

@article{Roychoudhury2025AgenticAF,
  title={Agentic AI for Software: thoughts from Software Engineering community},
  author={Abhik Roychoudhury},
  journal={ArXiv},
  year={2025},
  volume={abs/2508.17343},
}

@article{Sapkota2025VibeCV,
  title={Vibe Coding vs. Agentic Coding: Fundamentals and Practical Implications of Agentic AI},
  author={Ranjan Sapkota and Konstantinos I. Roumeliotis and Manoj Karkee},
  journal={ArXiv},
  year={2025},
  volume={abs/2505.19443},
}

%%
%% If your work has an appendix, this is the place to put it.
% \appendix

% \section{Research Methods}

% \subsection{Part One}

% Lorem ipsum dolor sit amet, consectetur adipiscing elit. Morbi
% malesuada, quam in pulvinar varius, metus nunc fermentum urna, id
% sollicitudin purus odio sit amet enim. Aliquam ullamcorper eu ipsum
% vel mollis. Curabitur quis dictum nisl. Phasellus vel semper risus, et
% lacinia dolor. Integer ultricies commodo sem nec semper.

% \subsection{Part Two}

% Etiam commodo feugiat nisl pulvinar pellentesque. Etiam auctor sodales
% ligula, non varius nibh pulvinar semper. Suspendisse nec lectus non
% ipsum convallis congue hendrerit vitae sapien. Donec at laoreet
% eros. Vivamus non purus placerat, scelerisque diam eu, cursus
% ante. Etiam aliquam tortor auctor efficitur mattis.

% \section{Online Resources}

% Nam id fermentum dui. Suspendisse sagittis tortor a nulla mollis, in
% pulvinar ex pretium. Sed interdum orci quis metus euismod, et sagittis
% enim maximus. Vestibulum gravida massa ut felis suscipit
% congue. Quisque mattis elit a risus ultrices commodo venenatis eget
% dui. Etiam sagittis eleifend elementum.

% Nam interdum magna at lectus dignissim, ac dignissim lorem
% rhoncus. Maecenas eu arcu ac neque placerat aliquam. Nunc pulvinar
% massa et mattis lacinia.

\end{document}